\documentclass[9pt,twocolumn,twoside]{opticajnl}
\journal{opticajournal} % use for journal or Optica Open submissions

\setboolean{shortarticle}{false}
\usepackage{float}
\usepackage{amsmath,esint}
\usepackage{indentfirst}
\usepackage[english]{babel}
\usepackage{fancyhdr}
\usepackage{lastpage}
\usepackage{tikz}
\usepackage{siunitx}
\usepackage{animate}
\usepackage{bm}
\usepackage{bbold}
\usepackage{wrapfig}
\usepackage{pdfpages}
\usepackage[justification=justified]{caption}
\usepackage{titlesec}
\usepackage{hyperref}
\usepackage{cleveref}
\usepackage{siunitx}
\usepackage{chngcntr}
\usepackage{svg}
\usepackage{comment}
\usepackage{mathtools}
\usepackage{stfloats}

\renewcommand{\thesubsection}{\thesection.\Alph{subsection}}

\usepackage{lineno}
\newcommand{\runin}[1]{%
    \vspace{1ex}           % Un petit espace vertical avant pour aérer
    \noindent\textbf{#1}   % Le titre en gras, sans alinéa
    \hspace{-0.5em}         % L'espace entre le titre et le texte
}

\DeclareSIUnit\bar{bar}
\DeclareSIUnit\gr{gr}

\DeclareCaptionFormat{myformat}{#1#2#3\hrulefill}
\title{High order harmonic Optical Reconstruction by delayed time-depeNdent Ellipticity perturbaTion: HORNET \\}
\author[1*]{Lucas Perez}
\author[1]{Corentin Picot}
\author[1]{Stefan Skupin}
\author[2]{Fabrice Catoire}
\author[1*]{Eric Constant}

\affil[1]{Université Lyon 1, CNRS, Institut Lumière Matière, UMR5306, F-69100, Villeurbanne, France}
\affil[2]{Université de Bordeaux, CNRS, CEA, Centre Laser Intense et Applications, Bordeaux, France}

\affil[*]{lucas.perez@univ-lyon1.fr, eric.constant@cnrs.fr}

\begin{abstract}
We present and demonstrate an all-optical \textit{in situ} method for temporal characterisation of high-order harmonics generated in gas with a linearly polarised intense infrared (IR) laser. 
By using a delayed cross-polarised replica of the IR pulse with a lower amplitude, we introduce a weak and time-dependent ellipticity in the IR driving field. Since the XUV emission is very sensitive to the ellipticity of the driving field, it is perturbed precisely at the times when ellipticity is significant. 
The detected emission is then reduced only if the unperturbed XUV emission is substantial at these times. Perturbed XUV spectra are recorded for several delays and exhibit changes both in amplitude and in their spectral content. An iterative ptychographic algorithm is used to retrieve the electric field of individual harmonics in a very robust way. Experimental measurements are in good agreement with simulations. 
The reconstructed temporal profiles show a steady evolution of the XUV pulse duration and chirp with the harmonic order. The highest harmonics exhibit a positive chirp and are emitted in the rising front of the IR driving pulse, while the lowest ones exhibit a negative chirp and are emitted throughout the entire pulse. 
\end{abstract}

\setboolean{displaycopyright}{false} % Do not include copyright or licensing information in submission.

\begin{document}

\maketitle

\section{Introduction} 

%    See \href{https://opg.optica.org/submit/templates/default.cfm}{Style Guide} and \href{https://opg.optica.org/submit/templates/default.cfm}{Manuscript Templates} pages for more details. 

%If you have a question while using this template on {Overleaf}, please use the help menu (``?'') on the top bar to search for help or ask us a question using our \href{https://www.overleaf.com/contact}{contact form}.

XUV attosecond pulse trains obtained via high-order harmonic generation (HHG) in gases \cite{ferrayMultipleharmonicConversion10641988b, mcphersonStudiesMultiphotonProduction1987b,corkumPlasmaPerspectiveStrong1993} are now commonly used to study ultrafast dynamics with possible sub-femtosecond resolution \cite{agostiniPhysicsAttosecondLight2004e, lepineMolecularApplicationsAttosecond2013, okinoNonlinearFourierTransformation2014, calegariAdvancesAttosecondScience2016, midorikawaProgressTabletopIsolated2022}. Since the discovery of the HHG process, many techniques have been developed to characterise temporally the XUV pulses. Most of them rely on the observation of a photoelectron spectrum as a function of the delay between the XUV pulse and a strong IR pulse as a reference \cite{paulObservationTrainAttosecond2001a, hentschelAttosecondMetrology2001a, mairesseFrequencyresolvedOpticalGating2005a, constantMethodsMeasurementDuration1997a, itataniAttosecondStreakCamera2002, sekikawaFrequencyResolvedOpticalGating2003,wangDirectReconstructionIsolated2023b}. 
% (ref Rabbitt, frog Crab, streaking, constant pra) \end{comment} 
These \textit{ex situ} techniques allow one to characterise the temporal profile of an XUV pulse at the position where photoionisation occurs. They are often difficult to implement as they require a photoelectron spectrometer and a high precision pump-probe setup. 

\textit{In situ} techniques that require simpler experimental setups have been developed to circumvent the above mentionned issues. They are based on characterising the XUV emission directly during the generation process to access the temporal profile of the XUV pulse exiting the generation medium. This optical characterisation is possible by introducing a weak perturbation during the generation process and observing its impact on the emitted light \cite{constantMethodsMeasurementDuration1997a}. Ideally, the perturbation acts at the single atom level, which makes it easier to model, understand, and control.
Initially a two-colour scheme with controllable \cite{dudovichMeasuringControllingBirth2006b, mauritssonSubcycleControlAttosecond2009} driver-perturbation delays was used to measure the temporal profile of an attosecond pulse in an attosecond pulse train assumed infinite and periodic. Later, the two-colour scheme also demonstrated his ability to determine the spatio-temporal structure of isolated attosecond pulses \cite{heAllopticalSpatiotemporalMetrology2022}. Recently, a single-colour approach was also developed to measure the temporal profile of isolated attosecond pulses \cite{yangAllopticalFrequencyresolvedOptical2020}. In all these cases, the polarisation of the main pulse and that of the perturbation pulse are identical and linear, which means that a slight change in intensity creates a so-called phase gate in the generation process.
Another single-colour scheme was also proposed, in which a constant ellipticity is introduced into the driving pulse to observe the timing of the harmonic emission on the attosecond scale. This approach relies on an amplitude gate and showed good agreement with the single-atom response \cite{dudovichAttosecondTemporalGating2006f}, which was simulated using a semi classical approach, again assuming an infinite periodic attosecond pulse train. So far, \textit{in situ} methods have mainly focused on characterising infinitely long attosecond pulse trains or isolated attosecond pulses. In reality, XUV pulses are often between these two limits and it is very important to characterise the temporal profile of each harmonic to increase the measurement precision. This characterisation was successfully carried out using a SPIDER approach \cite{mairesseHighHarmonicXUV2005}, although its implementation remains challenging because it requires XUV spectral shearing interferometry. %have been used to characterise temporally attosecond pulses emitted by HHG. 
Furthermore, many applications such as coherent diffraction imaging \cite{ravasioSingleShotDiffractiveImaging2009} or time and angle resolved photoemission spectroscopy (ARPES) \cite{eichTimeAngleresolvedPhotoemission2014} require the use of only one harmonic of the comb, and it remains crucial to characterise the femtosecond temporal profile of each harmonic. 

In this work, we present a new \textit{in situ} method, that we called HORNET, for measuring the temporal profiles of individual harmonics. This technique relies on the well-known strong dependence of HHG on the ellipticity of the driving field \cite{budilInfluenceEllipticityHarmonic1993b, mollerDependenceHighorderharmonicgenerationYield2012c}. More specifically, this method is based on the introduction of time-dependent ellipticity in the IR driving field during the generation process. The time-dependent ellipticity is created by adding a perpendicularly polarised weak perturbation pulse that co-propagates with the strong main pulse. Using a perturbation pulse with a polarisation perpendicular to the polarisation of the main pulse has several key advantages. It creates an amplitude gate whose impact can be measured directly \cite{budilInfluenceEllipticityHarmonic1993b} without relying on models. It also allows for strong XUV signal modification without changing the driving field amplitude significantly and thereby without affecting the macroscopic generation conditions (ionisation yield, phase matching, etc.). For example, if a perturbation pulse with 1\% of the energy of the main pulse is used, the field amplitude changes by only 0.5\% when the polarisations are perpendicular to each other, but by $\pm$ 10\% when the polarisations are parallel. In the latter case, macroscopic parameters may be affected \cite{yanAmplitudeModulationEffect2025a}, which are difficult to account for without complex models. % This effectively disentangle perturbing the single atom response and the collective response that is often a limitation with phase gates.

Delaying the perturbation pulse compared to the main pulse modulates the total IR field ellipticity in time, and thereby moves an ellipticity perturbation gate within the driving pulse. %\cite{corkumSubfemtosecondPulses1994, mollerDe+pendenceHighorderharmonicgenerationYield2012c, mauritssonSubcycleControlAttosecond2009}
This attenuates high-order harmonic emission in a well-controlled part of the driving pulse. It reduces the measured XUV signal only if unperturbed XUV emission occurs in this specific part of the pulse, and it has no effect if the unperturbed emission is zero. Thus, delaying the perturbation provides access to the temporal profile of the unperturbed XUV emission. 

The perturbation pulse is a weak replica of the main pulse with perpendicular polarisation.
%obtained through a polarisation dependent interferometer, as detailed in the experimental Section~\ref{sec:exp-set-up}.
Using a replica of the main pulse as the perturbation results in a much simpler setup than in the two-colour scheme mentioned above, as no frequency doubling crystal is required. %In addition, the amplitude of the main field is not significantly changed by the perturbation, leaving the macroscopic conditions unchanged. 
In addition, it implies that both pulses will propagate in the same way inside the generating medium. %, at least as far as linear propagation effects are concerned. 
XUV recorded spectra change with the delay in terms of both their amplitude and spectral content, and this evolution contains all the information about the temporal profile of the unperturbed XUV pulse. A ptychographic phase retrieval technique \cite{lucchiniPtychographicReconstructionAttosecond2015} allows to extract the amplitudes and spectral phases of the harmonics from the dependence of the perturbed harmonic spectra on the delay between main pulse and perturbation. 
We demonstrate a full reconstruction of individual harmonic temporal electric fields -- both intensity and phase --  first through a simple semi-analytical approach using known complex XUV pulse profiles and later experimentally to characterise high-order harmonics generated in a gas cell filled with argon. Our experimental results show a steady evolution of the harmonic pulse durations and chirps.
%the average emission time of each harmonic and the harmonic envelop temporal profile can be reconstructed with a simple approach. 

%Introduction: many existing characterization techniques relying on electron spectroscopy. Important to develop optical techniques (and some are appearing currently). In situ and ex situ characterization: here optical in situ for characterization of the harmonic envelop.

\section{Materials and methods} 

The proposed technique %, named HORNET for "High order harmonic Optical Reconstruction by delayed time-depeNdent Ellipticity perturbaTion", 
%is based on the well-known strong ellipticity dependence of HHG in gases \cite{budilInfluenceEllipticityHarmonic1993b}. It 
relies on introducing a time-dependent ellipticity perturbation in the IR driving pulse and analysing the changes of the XUV perturbed spectra with the position of the perturbation gate within the pulse. The measurement is performed directly in the generating medium (\textit{in situ} method) and spectra are measured with an XUV spectrometer. % optically.

\subsection{Principle of ellipticity perturbation}  \label{sec:ellipt_theory}

Time-dependent ellipticity is introduced through a polarisation-splitting interferometer that separates a linearly polarised IR pulse into two delayed cross-polarised pulses: a main driving pulse and a weak perturbation pulse. %Great care is taken to ensure that the two pulses have no chirp in the generating medium.

The key idea lies in the fact that introducing ellipticity at times when the unperturbed XUV intensity is significant results in a decreasing XUV signal on the spectrometer. % if the  the ellipticity perturbation only affects the XUV signal when it is located at a position where the unperturbed signal is significant.
Otherwise, it has no effect on the measured signal. Therefore, delaying the ellipticity perturbation probes the XUV temporal profile.
When the delay is negative (positive), the perturbation will affect the XUV light generated in the rising (falling) front of the IR pulse. % Furthermore, a constant ellipticity provides the ellipticity sensitivity of the harmonics, which happens near the zero delay.

We define $\delta$ as the delay between the envelopes of the main pulse $\tilde E^{\textrm{main}}_{\textrm{IR}}(t)$ and the perturbation pulse $\tilde E^{\textrm{p}}_{\textrm{IR}}(t,\delta)$. A small ellipticity (around 0.1) is typically sufficient to decrease the XUV intensity by a factor 2 \cite{budilInfluenceEllipticityHarmonic1993b}. Therefore, the intensity of the perturbation pulse is set to approximately 1\% of the intensity of the main pulse: $\tilde E^{\textrm{p}}_{\textrm{IR}}(t,\delta)\approx 0.1 \tilde E^{\textrm{main}}_{\textrm{IR}}(t-\delta)$. With such a low intensity of the perturbation, the physical properties that define the unperturbed XUV pulse temporal profile (intensity in the generating medium, beam spatial profile, ionisation rate, phase matching) are not modified.

\begin{figure}[t]
    \centering
    \includegraphics[width=1\linewidth]{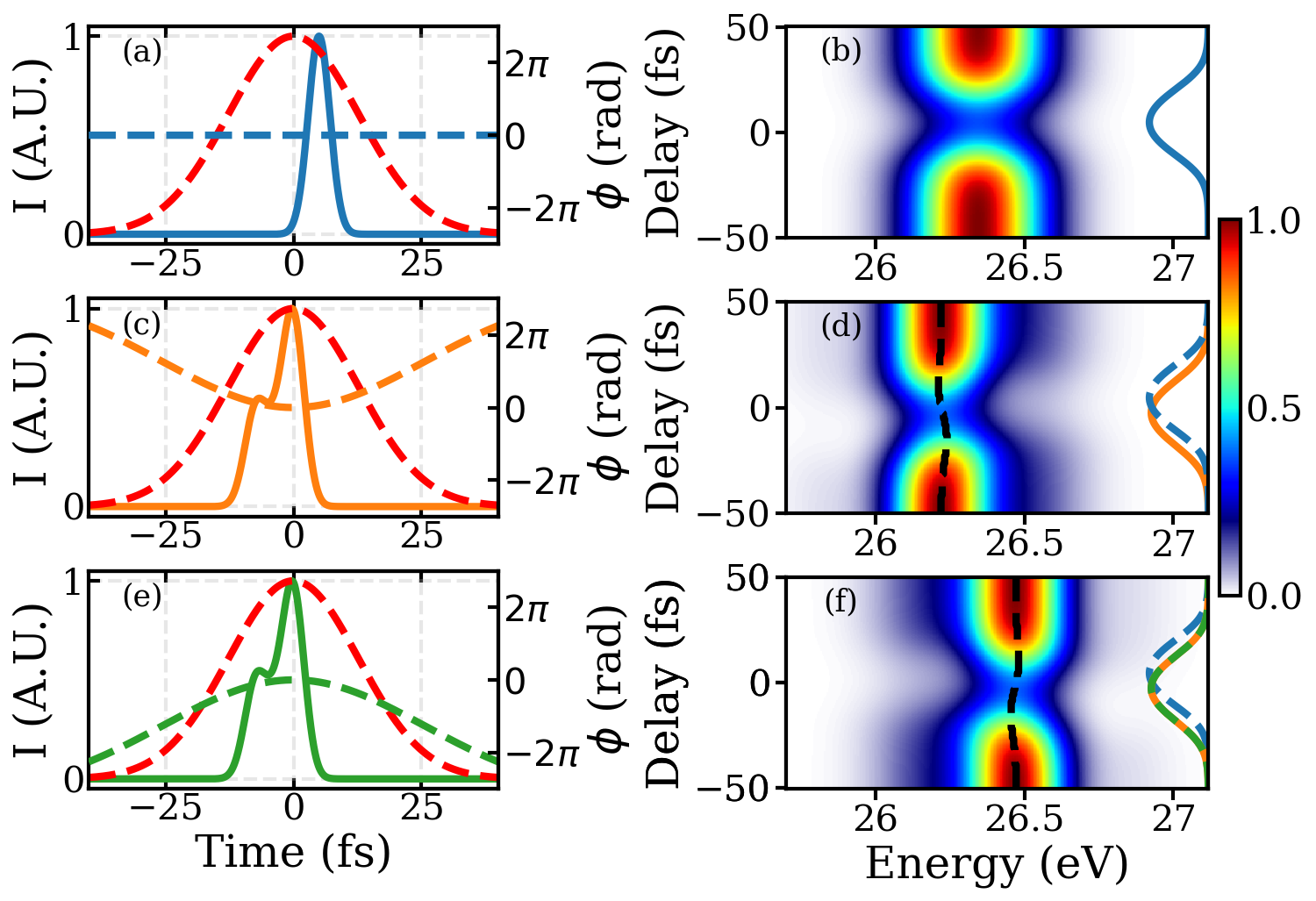}
    \caption{Ellipticity perturbation traces obtained from Eq.~(\ref{eq:spectrum}) with envelope ellipticity, see text for details. Left: three XUV intensity (solid lines) and phase (dashed lines) profiles (blue (a): 5~fs FWHM Gaussian profile centered at +5~fs; orange (c) and green (e): two 5~fs FWHM Gaussian profiles spaced by 7.5~fs with a $1:2$ intensity ratio) and the normalised intensity of the main IR pulse (32~fs FWHM) in dashed red line. The phases are indicated in dashed colored lines. Right: corresponding ellipticity perturbation traces obtained in the spectral domain (b, d, f). Blue, orange and green curves are the integrated signal obtained by summing over all frequencies and called lower envelopes. Blue and orange curves are also shown as dashed lines for comparison. %The blue curve is also shown as a dashed line in (d, f) for comparison. 
    Dotted black curves in (d, f) follow the XUV central frequency for each delay.}
    \label{fig:simu_maps}
\end{figure}

%It is known that the HHG efficiency decays with the ellipticity of the IR driving pulse \cite{budilInfluenceEllipticityHarmonic1993b}. 
The HHG efficiency decays with the ellipticity of the IR driving field and for small constant ellipticities, this decay can be approximated by a Gaussian function \cite{mollerDependenceHighorderharmonicgenerationYield2012c, heSingleAttosecondPulse2007, antoineTheoryHighorderHarmonic1996}. We extrapolate this expression to the case where the ellipticity is time-dependent. The validity of this assumption has been established by the so-called polarisation gating approach \cite{corkumSubfemtosecondPulses1994, solaControllingAttosecondElectron2006, tcherbakoffTimegatedHighorderHarmonic2003, kovacevTemporalConfinementHarmonic2003, gilbertsonIsolatedAttosecondPulse2010} where the ellipticity of the driving pulse significantly changes over one optical period to confine the XUV emission. Here, the ellipticity evolves more slowly and the ellipticity is small during the XUV emission, making the assumption even more reliable. %since the perturbation pulse and the main pulse have the same duration. 
We approximate the perturbed XUV spectrum of the $q^{\text{th}}$ harmonic by
\begin{equation} \label{eq:spectrum}
    S_{\text{q}}(\omega,\delta) \approx \left| \int_{-\infty}^{+\infty} \mathrm{d}t \,E_{\text{q}}(t) \exp(-\beta_{\text{q}} \, \varepsilon^2(t,\delta)) \exp(-\textrm{i} \omega t) \right|^2,
\end{equation}
where $E_{\text{q}}(t)$ is the unperturbed XUV complex-valued electric field to be retrieved, $\beta_{\text{q}} = -\ln(2)/2\varepsilon^2_{\text{thr}}(q)$, and $\varepsilon_{\text{thr}}(q)$ is the threshold ellipticity reducing the XUV signal of the $q^{\text{th}}$ harmonic by a factor 2 (for constant ellipticity). This threshold parameter depends on the gas used, the fundamental wavelength and the harmonic order $q$.

The time-dependent ellipticity $\varepsilon(t,\delta)$ of the driving pulse introduced by the interferometer depends on the ratio $\tilde E^{\textrm{p}}_{\textrm{IR}}(t,\delta)/\tilde E^{\textrm{main}}_{\textrm{IR}}(t)$ and the dephasing $\delta\varphi$ between the two electric fields. As the delay varies, the dephasing also varies between $\delta \varphi=0$ -- the ellipticity is zero throughout the pulse which provides a reference unperturbed XUV spectrum --  and $\delta\varphi=\pi/2$ -- the ellipticity is maximum and the XUV signal is locally minimum. This leads to oscillations in the XUV signal with a period $T_0/2$, where $T_0$ is the optical period of the central frequency of the IR driving pulse. The oscillation period thus provides a self-reference time scale. % These rapid oscillations of the signal provides a self reference for the delay. 
One can artificially remove these oscillations by considering the envelope ellipticity
\begin{equation} \label{eq:envelope_ellipticity}
\tilde\varepsilon(t,\delta) = \tilde E^{\textrm{p}}_{\textrm{IR}}(t,\delta)/\tilde E^{\textrm{main}}_{\textrm{IR}}(t),
\end{equation}
which is equivalent to assuming a $\delta\varphi=\pi/2$ dephasing for all delays. This can be done by selecting only the spectra obtained for $\pi/2$ dephasing and interpolating between them. Note that evaluating Eq.~(\ref{eq:spectrum}) at zero delay, which corresponds to a constant ellipticity in time, can be used to extract the value of $\varepsilon_{\text{thr}}(q)$ from experimental measurements. %toujours pas hyper clair 

\begin{figure*}[t]
    \centering
    \includegraphics[width=0.9\linewidth]{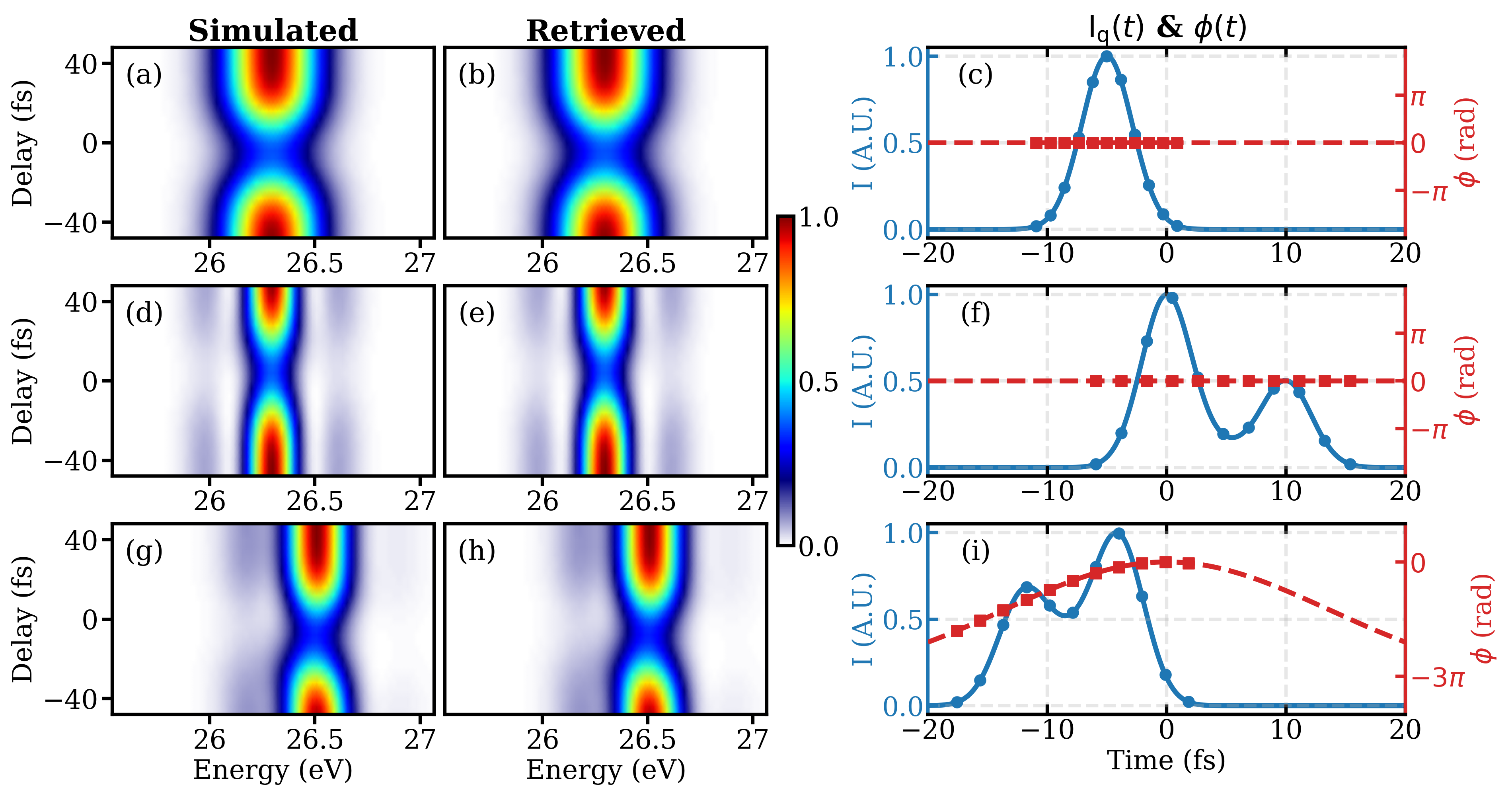}
    \caption{Reconstruction of XUV temporal profiles from ellipticity perturbation traces: intensities in blue solid lines and phases in red dashed lines (c, f, i). Corresponding simulated ellipticity perturbation traces from Eq.~(\ref{eq:spectrum}) (a, d, g) and retrieved traces (b, e, h) by the ptychographic algorithm. Retrieved temporal profiles are indicated by circle markers for intensities and square markers for phases.}
    \label{fig:simulated_reconstructions}
\end{figure*}

In Fig.~\ref{fig:simu_maps}, we evaluate Eq.~(\ref{eq:spectrum}) to obtain simulated ellipticity perturbation traces for three known XUV temporal profiles. An IR Gaussian driver pulse centred at 800~nm with a full-width at half maximum (FWHM) of 32~fs and no chirp is used (Fig.~\ref{fig:simu_maps}, red dashed line). The envelope time-dependent ellipticity is calculated with a perturbation energy equal to 2\% of the main pulse energy using Eq.~(\ref{eq:envelope_ellipticity}). The known XUV temporal profiles are shown together with the driver pulse envelope in the left column (a, c, e), and  the temporal phases are depicted with same colour dashed lines. The XUV profile in (a) is considered Fourier-transform limited, while the other two profiles have an intensity-dependent dipole phase $\phi(t) = \alpha I_{\text{IR}}(t)$ to mimic the femtochirp in the harmonic field arising from quantum contributions to the generation process \cite{lewensteinPhaseAtomicPolarization1995,gaardeSpatiotemporalSeparationHigh1999a}, where $\alpha<0$ in (c) and $\alpha>0$ in (e). The chosen values of $\alpha = \qty{5e-14}{W^{-1} \, cm^2}$ and $I_{\textrm{max}}=\qty{2e14}{W \, cm^{-2}}$ agree with calculations in argon (Ar) for the short trajectory \cite{heSpatialSpectralProperties2009, catoireComplexStructureSpatially2016a}. % but were increased to be clearly visible on the traces. 
%The time-dependent envelope ellipticity Eq.~\ref{eq:envelope_ellipticity} was calculated for each delay taking into account the full propagation in the quartz and fused silica plates that make up the polarisation dependent interferometer, cf.~\cref{sec:exp-set-up}. 
The resulting traces are presented in the second column for the three XUV profiles (Fig.~\ref{fig:simu_maps} b, d, f) and are clearly different. The coloured curve on the right represents the spectrally integrated signal and is called the lower envelope in what follows. %The black curve represents the mean spectra over all delays while the coloured curve represents the spectrally integrated signal called lower envelope. 
When the delay becomes much larger than the IR pulse duration, the perturbation does not introduce any significant ellipticity during HHG and the spectra converge to the unperturbed spectrum. %, to which the traces are normalised. 
The traces exhibit clearly different behaviours, strongly depending on the XUV profile, and therefore contain useful information about it. The spectrum at large delays directly provides the spectrum of the unperturbed XUV pulse, while the changes in the spectrum, when the emission is perturbed, give information on the XUV chirp and the location of the XUV pulse relative to the IR pulse. 
In panel (b), the blue lower envelope is shifted towards positive delays, since the emission in (a) occurs mainly for positive times. On the other hand, the orange lower envelope in panel (d) is shifted towards negative delays, since the emission in (c) occurs mainly for negative times. This removes any ambiguity on the time axis direction. 
The evolution of the instantaneous XUV frequency is also encoded in the traces, as depicted in panels (d) and (f). Indeed, since the emission occurs mainly for positive delays, the spectra are shifted towards red (d) and blue (f) frequencies because of the XUV femtochirp. In addition, the lower envelopes in (d) and (f) are identical (green and orange curves), illustrating that they are not sensitive to the temporal phase of the XUV pulse. However, the XUV spectra depend on this phase, and when the XUV pulse is chirped, the central frequency changes with the delay, exhibiting a "s-shape" (black dashed lines), unveiling the evolution of the instantaneous frequency inside the pulse. % The temporal phase information are therefore inscribed in the traces. % can also be observed (row 3 of \cref{fig:simulated_reconstructions}) along the delay, providing information on temporal phase. %Note that at zero delay the ellipticity is constant equals to $\tan(\theta)$. This provides information on $\varepsilon_{\text{thr}}(q)$ for each harmonic. 

\subsection{Ptychographic algorithm for the reconstruction of the XUV temporal profile}

Even if the physical information is directly visible on the trace, it is not straightforward to extract the unperturbed XUV electric field. Iterative algorithms can be used for this purpose. Eq.~(\ref{eq:spectrum}) defining the ellipticity perturbation traces is very similar to the equations used in FROG-like techniques \cite{kaneCharacterizationArbitraryFemtosecond1993a} with a perturbation ellipticity gate $\exp(-\beta \, \varepsilon^2(t,\delta))$ scanned in delay. The main difference lies in the fact that the gate itself changes with the delay. This is not a fundamental difference, provided that the time-dependent ellipticity is precisely known for each delay. This will be discussed in Section~\ref{sec:robustness}. Several algorithms such as the Principal Component Generalised Projects Algorithm (PCGPA) \cite{kaneRecentProgressRealtime1999a} can be used to analyse these types of traces. However, ptychographic algorithms have been demonstrated to further improve convergence and robustness to noise \cite{lucchiniPtychographicReconstructionAttosecond2015, sidorenkoPtychographicReconstructionAlgorithm2016a}. Based on these previous works, we have developed a method to reconstruct XUV electric fields from ellipticity perturbation traces. Later in Section~\ref{sec:exp} these traces will come from experimental measurements, but here we first test the method with simulated traces as described in Section~\ref{sec:ellipt_theory}, for which the unperturbed XUV field is known.

The proposed reconstruction algorithm works as follows. First, we use the IR temporal profile to calculate the ellipticity profile $\tilde\varepsilon(t,\delta_j)$ for some discrete delays $\delta_j$ ($j = [1,J]$) in the range of interest. For simplicity, we use the envelope ellipticity Eq.~(\ref{eq:envelope_ellipticity}) for both the simulation and the reconstruction of the traces. %, i.e., a dephasing of $\delta\varphi=\pi/2$ between the envelopes of the main IR pulse and the perturbation pulse is assumed for all delays. 
Second, an initial guess for the unknown XUV field is considered. A reasonable choice which works well in practice is the inverse Fourier transform of the unperturbed XUV spectral amplitude, i.e., the shortest pulse compatible with the unperturbed spectrum. %By contrast, choosing a random initial guess can easily lead to divergence of the algorithm and failed reconstruction. 
Third, we start a loop of $N$ iterations where each iteration is itself an internal loop of $J$ iterations running over the delays. At each iteration step, the algorithm compares the perturbed spectrum of the current guess to the perturbed spectrum at $\delta_j$ and updates the XUV profile.   
%where we compute the perturbed spectrum of the current guess for the delay $\delta_j$, compare its modulus with the trace at the same delay, and update the XUV profile as described below. 
%This process is repeated $N$ times. 
We now describe the iteration step to obtain the $j^{\textrm{th}}$ modification of the unperturbed XUV pulse, i.e., obtain the XUV field $E_{j+1}(t)$ from $E_j(t)$ and the simulated (or measured) spectrum at delay $\delta_j$. For better readability, we omit the index $q$, since only one harmonic is considered for the analysis. We start by calculating the perturbed electric field $\psi_j(t)$ that would be generated for the delay $\delta_j$ with the field $E_j(t)$:
\begin{equation}
    \psi_j(t) = E_j(t) \exp(-\beta \, \tilde\varepsilon^2(t,\delta_j)).
\end{equation}
This perturbed field is then Fourier transformed to the spectral domain, where its modulus is replaced by the amplitude of the simulated (or measured) spectrum at this delay and the phase is kept unchanged. The resulting updated spectrum is then Fourier transformed back to the temporal domain:
\begin{equation}
    \psi'_j(t) = \mathcal{F}^{-1}\left[\sqrt{S(\omega,\delta_j)} \dfrac{\mathcal{F}[\psi_j(t)]}{|\mathcal{F}[\psi_j(t)]|}\right],
\end{equation}
where $\mathcal{F}$ denotes the Fourier transform. Finally, the estimated XUV field is updated using a weight function based on the perturbation ellipticity gate:
\begin{equation}
    E_{j+1}(t) = E_j(t) + \gamma \, \exp(-\beta \, \tilde\varepsilon^2(t,\delta_j)) \, (\psi'_j(t) - \psi_j(t)),
\end{equation}
where $0<\gamma<1$ is a parameter that controls the strength of the correction. A rather small $\gamma$ is needed to keep track of the updates made at each iteration and converge to a solution that accounts for all the delays considered. We find that $\gamma=0.2$ provides a good compromise between speed of convergence and accuracy for the reconstructions performed in this work. This loop is repeated $N$ times or until some convergence criterion is met.

\subsection{Validation of the reconstruction}

We first validate our algorithm by retrieving known XUV electric fields. We simulate the traces generated with a Gaussian pulse centred at 800 nm, FWHM = 32 fs and a maximal intensity $I_{\textrm{max}}=\qty{2e14}{W \, cm^{-2}}$ for three temporal profiles of harmonic 17. The delay is scanned in the range of -50~fs to +50~fs with a step of 4~fs. The results are presented in Fig.~\ref{fig:simulated_reconstructions}. The first column (a, d, g) displays the simulated traces corresponding to the temporal profiles shown in the third column (c, f, i). The simulated profiles are presented with blue solid lines for intensities and red dashed lines for phases. 
In the first row, we simulate a 5~fs FWHM Gaussian pulse centred at -5~fs. In the second row, we simulate two 5~fs FWHM Gaussian pulses centred at 0~fs and +10~fs with a ratio of one to two in intensity and a flat temporal phase. In the third row, we simulate two 5~fs FWHM Gaussian pulses centred at -4~fs and -12~fs with a ratio of one to two-thirds in intensity with temporal phase $\phi(t)=\alpha I_{\textrm{IR}}(t)$ ($\alpha = \qty{5e-14}{W^{-1} \, cm^2}$). 
The second column (b, e, h) displays the retrieved traces obtained by the ptychographic algorithm. The retrieved and simulated traces are almost identical (RMSE < 0.001\%). % After a thousand iterations, the error goes down to $10^{-7}$.
The markers (circles for intensities and squares for phases) depicted in the third column indicate the retrieved profiles. They are only shown in a limited temporal range for the sake of readability but are retrieved for all times. %not sure of this formulation
In all cases, the retrieved pulse matches the simulated pulse (lines) for both intensity profile and temporal phase. The intensity profile converges in a few tens of iterations to the mean emission time since this information is strongly encoded in the lower envelope, while the complete reconstruction of the electric field may typically require a thousand iterations to converge. Note that this method is unsensitive to the carrier-envelope phase (CEP), so a constant phase offset exists between the simulated and retrieved electric fields, with no impact on the intensity profiles. 

\begin{figure}[t]
    \centering
    \includegraphics[width=1\linewidth]{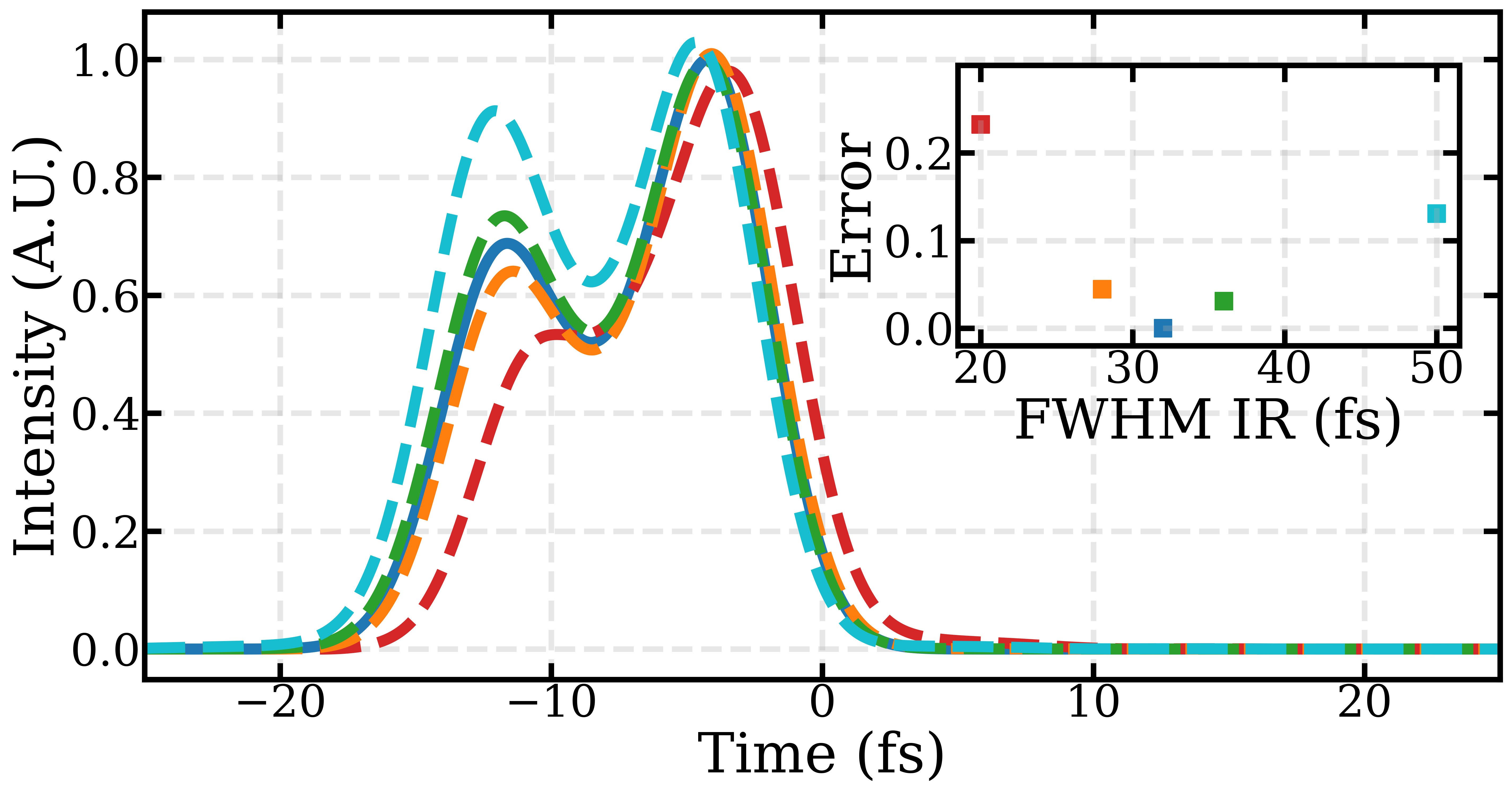}
    \caption{XUV intensity profiles (dashed lines) reconstructed using Gaussian IR profiles with several FWHM durations (red: 20~fs, orange: 28~fs, green: 36~fs and cyan: 50~fs). The original simulated profile is the blue solid curve (32~fs). The inset shows the error of the reconstructed profiles calculated using Eq.~(\ref{eq:error}).}
    \label{fig:robustness_IR}
\end{figure}

\subsection{Robustness} \label{sec:robustness}

Experimental measurements can suffer from some uncertainties that can affect the quality of the reconstruction. In this section, we investigate and quantify the impact of these uncertainties numerically. %using simulations.

As discussed in Section~\ref{sec:ellipt_theory}, determining the time-dependent ellipticity profile requires high-precision measurement of the IR field. Since this measurement can suffer from noise and uncertainties, especially on the FWHM duration, we test the robustness of our algorithm against deviations in the duration of the IR driving pulse and thus against an incorrect ellipticity profile for reconstruction. We start with the XUV temporal profile shown in Fig.~\ref{fig:simulated_reconstructions}~(i) and the corresponding trace (g), obtained with a Gaussian IR profile of 32~fs FWHM. Then, we generate time-dependent envelope ellipticities $\tilde\varepsilon(t,\delta)$ using Gaussian IR profiles with durations of 20, 28, 36 and 50~fs FWHM and reconstruct the corresponding XUV temporal profiles with our ptychographic algorithm. To quantify the error of the reconstructed XUV profiles, we use the following expression, which is not sensitive to the CEP of the electric field \cite{kohlImprovedSuccessRate2012}:
\begin{equation} \label{eq:error}
    \theta(E_{\text{q}},\hat{E}_{\text{q}}) = \arccos \left(\dfrac{| \langle E_{\text{q}}(t),\hat{E}_{\text{q}}(t) \rangle |}{\sqrt{\langle E_{\text{q}}(t),E_{\text{q}}(t) \rangle\langle \hat{E}_{\text{q}}(t),\hat{E}_{\text{q}}(t) \rangle}} \right),
\end{equation}
where $\hat{E}_{\text{q}}(t)$ and $E_{\text{q}}(t)$ are, respectively, the target and retrieved electric fields. $\langle f(t),g(t)\rangle = \int \mathrm{d}t \, f(t) g^*(t)$ is the inner product between the two fields, where $^*$ denotes the complex conjugate.

The results of the reconstruction are presented in Fig.~\ref{fig:robustness_IR} with the XUV intensity and temporal phase profiles retrieved for each IR duration, and the inset gives the error associated with each profile. In all cases, the algorithm converged to a physically meaningful result. The target profile is plotted as a blue solid line in Fig.~\ref{fig:robustness_IR} while the retrieved profiles with the wrong IR durations of 20, 28, 36 and 50~fs FWHM duration are plotted as red, orange, green, and cyan dashed lines, respectively. The associated errors are, respectively, 0.22, 0.05, 0, 0.03, and 0.12 (rounded to the second decimal). These errors and the corresponding retrieved XUV intensity profiles indicate that ptychographic reconstruction works for errors of up to 10\% in IR duration. %A perfect characterisation of the IR pulse is therefore not crucial as long as it remains within reasonable limits and the retrieval remains constrained around the initial profile without rapidly diverging for both intensity and phase.

\begin{figure*}[t]
    \centering
    \includegraphics[width=0.9\linewidth]{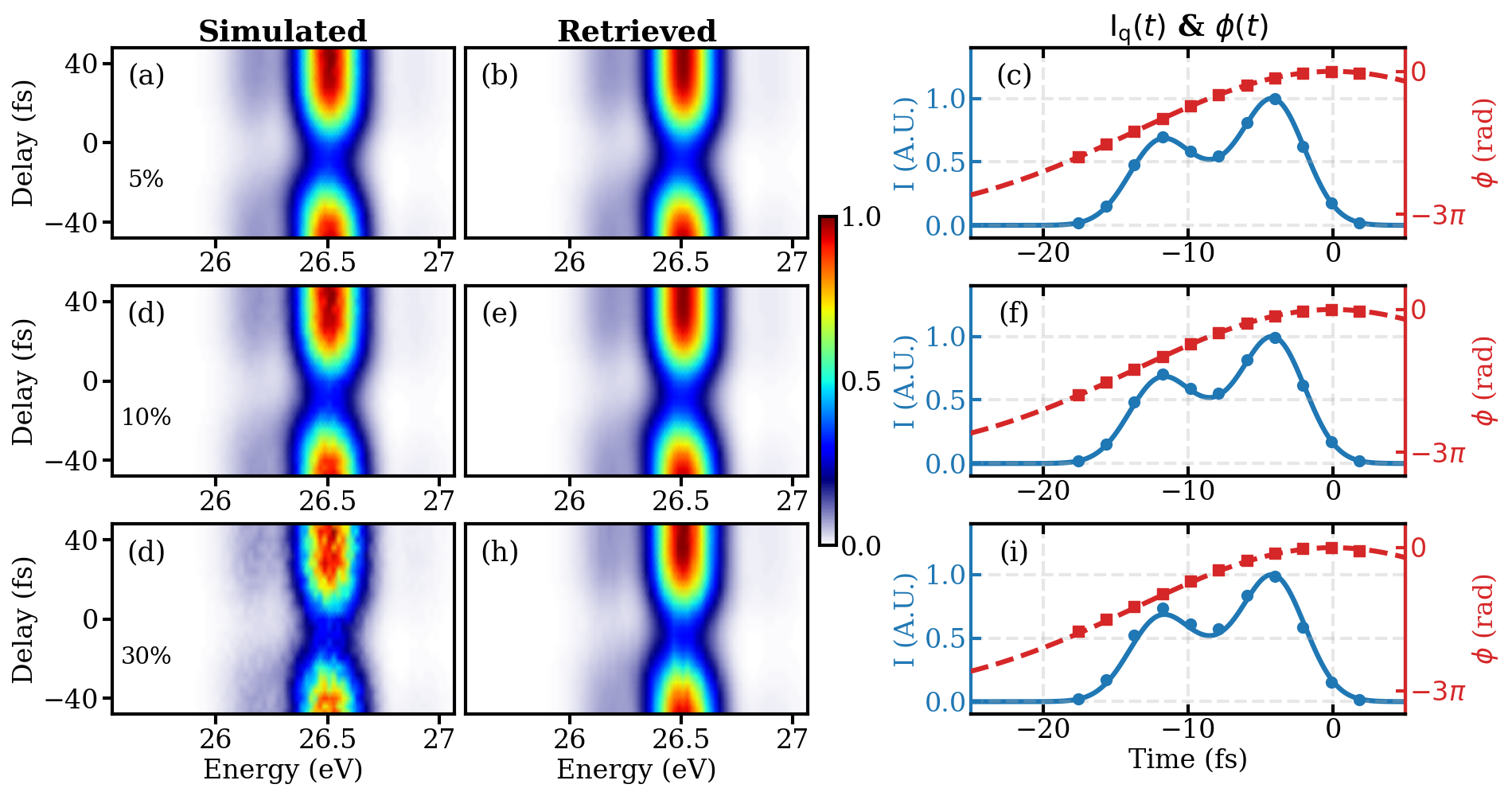}
    \caption{Column 1: simulated ellipticity perturbation traces with noise level of 5\% (a), 10\% (d) and 30\% (g). Column 2: retrieved traces. Column 3: simulated reference intensity (blue solid line), phase (red dashed line) and corresponding retrieved profiles (markers).}
    \label{fig:robustness_noise}
\end{figure*}

Noise in measured data is another frequent experimental artefact that can be problematic for reliable reconstruction. To quantify the sensitivity of the reconstruction procedure to noise in the ellipticity perturbation trace, we still consider the temporal XUV profile presented in Fig.~\ref{fig:simulated_reconstructions}~(i) and the corresponding trace (g), and we add random numbers between -0.5 and 0.5 multiplied by the signal and a noise level percentage. This random noise models the fluctuations that may occur during the measurements such as fluctuations in the laser peak intensity or in detector response. The traces with added noise levels of 5\%, 10\% and 30\% are presented in Fig.~\ref{fig:robustness_noise}~(a), (d), and (g), respectively. %From a noise level of 10\%, the distortion of the trace is clearly visible. 
The retrieved traces shown in (b), (e), and (h) are considerably smoother and contain the core information of the original noise-less trace. This manifests itself in the retrieved temporal profiles shown with circle markers in (c), (f), and (i) with corresponding errors (Eq.~(\ref{eq:error})) of 0.01, 0.02, and 0.05, respectively. The original profile is plotted as a blue solid line for the intensity and a red dashed line for the phase. The low values for the errors demonstrate the robustness of the algorithm in extracting XUV electric fields even in the presence of noise in the signal. 

\section{Experimental results}\label{sec:exp}

In what follows, we use the proposed method to experimentally measure ellipticity perturbation traces and apply our algorithm to extract the XUV temporal profiles of individual harmonics. 

\subsection{Experimental set-up} \label{sec:exp-set-up}

%We do not precise the spatial range of integration

\begin{figure}[t]
    \centering
    \includegraphics[width=1\linewidth]{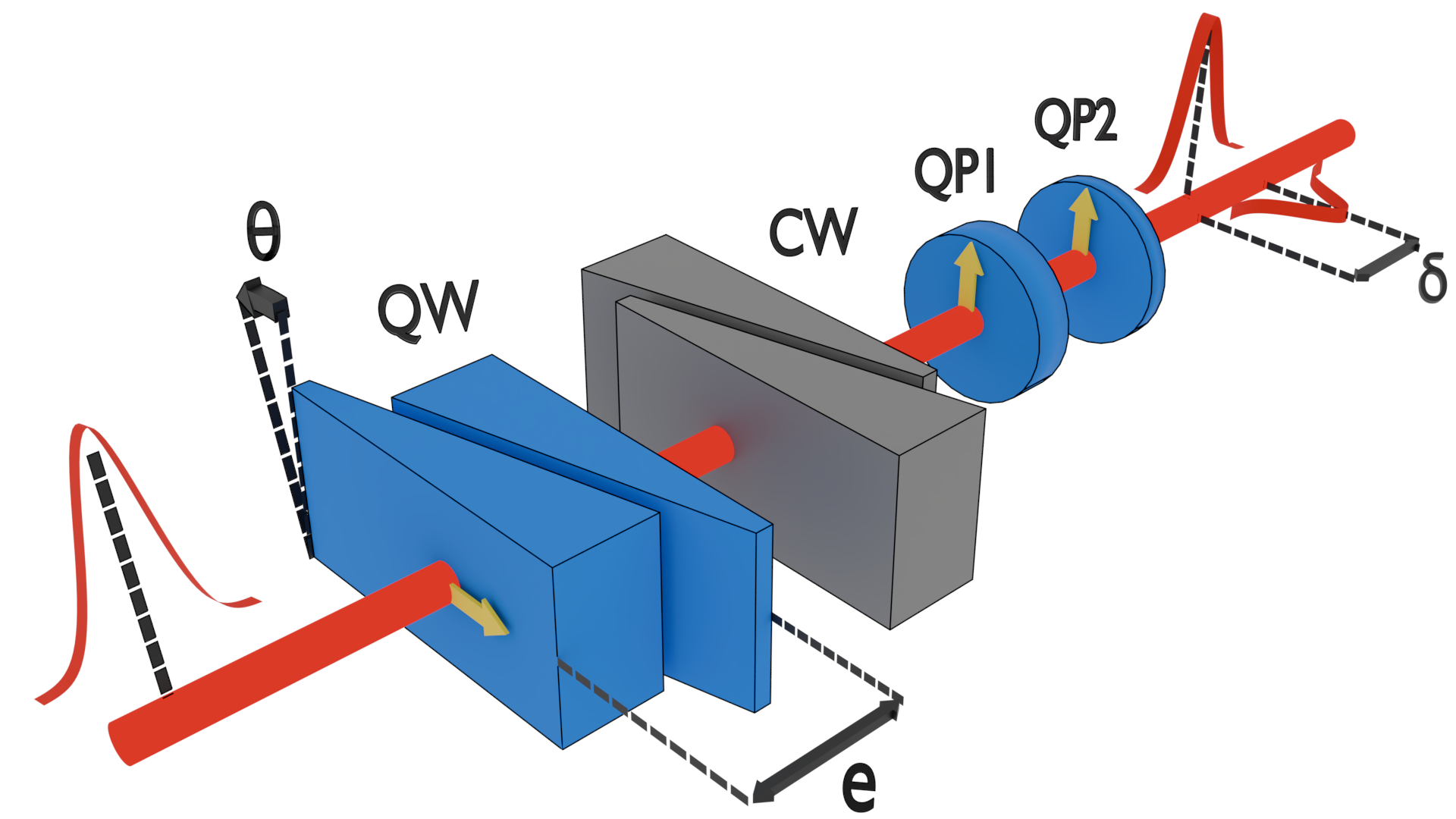}
    \caption{Scheme of the set-up used to create a cross-polarised replica of the main driving pulse pulse with a tunable delay. The quartz birefringent wedges (QW) separate the input pulse in two cross-polarised replicas with amplitude ratio given by the angle $\theta$. The extraordinary axes of the wedges are horizontal, as indicated by the yellow arrow and the input polarisation is close to the vertical (angle $\theta$). The delay $\delta$ between the two replicas is controlled by moving the wedges and changing the total thickness $e$. The non-birefringent compensating wedges (CW), the quartz waveplate 1 (QP1) and the quartz waveplate 2 (QP2) are depicted as well. QP1 has its extraordinary axis (yellow arrow) perpendicular to the one of the wedges. The two delayed pulses are then sent to a vaccum chamber and focused into a 6~mm thick cell filled with Ar to generate high-order harmonics.}
    \label{fig:Wedges}
\end{figure}

The experimental setup consists of a Ti:sapphire laser system delivering $\qty{32}{\fs}$ pulses at $\qty{800}{nm}$ with a repetition rate of $\qty{5}{\kilo\Hz}$ and $\qty{2}{\milli\J}$ energy per pulse. The pulse propagates through a polarisation-splitting interferometer, which consists of two birefringent quartz wedges that separate the incoming pulse into two: a main pulse and a delayed cross-polarised perturbation pulse. These wedges are depicted by QW in Fig.~\ref{fig:Wedges} and have the same horizontal optical axis as indicated by the yellow arrow. They are thus equivalent to a birefringent quartz plate with a controllable thickness. 
The input pulse is linearly polarised in a direction that makes an angle $\theta$ with the vertical. This angle defines the ratio between the main pulse and the perturbation pulse field amplitudes. It must remain below $\qty{10}{\degree}$ to ensure that the intensity of the perturbation pulse is less than 2\% of that of the main pulse, and thus the generation parameters are not affected, as discussed in Section~\ref{sec:ellipt_theory}. In our experiments, we used $\theta = \qty{8}{\degree}$.  % Doing so, the calcite plate separates the laser pulse in a main pulse and a small pulse delayed by $\qty{500}{\fs}$ thus ensuring a clean linear polarisation rotated by $\qty{8}{\degree}$. 

Using wedges has several advantages, such as 1) a straightforward separation of the incoming pulse into main pulse and perturbation pulse; 2) the collinear propagation of the two orthogonally polarised pulses ensures identical propagation ranges in the glass, making it easier to compensate for dispersion, and no additional alignment for spatial overlap in the interaction volume is required; 3) the wedge angle is small (measured as $\qty{1}{\degree}$), so that a large variation of the translation stage on which the wedges are mounted causes a small change of the optical thickness, thus enabling sub-femtosecond control of the delay between the two pulses. %In our setup, the two wedges move at the same time in opposite directions, doubling the range of accessible delays. 
Regarding the dimensions of the beam (waist radius $w = \qty{4.2}{\mm}$) and the $\qty{50}{\mm}$-long wedges, the interferometer has a maximum delay range of $\delta=\qty{40}{\fs}$. We measured the interferometer static delay stability to be $\sigma=\qty{1.8}{\as}$ over two hours, which is comparable to other similar measurements \cite{jansenSpatiallyResolvedFourier2016}. % without obstructing the beam. 
Three limitations arise with this setup, and we present our solutions to overcome them. 
First, the perturbation is only positively delayed since the optical axis of the quartz is a slow axis. To scan both negative and positive delays, we add a quartz waveplate (QP1 in Fig.~\ref{fig:Wedges}) with its optical axis orthogonal to the one of the wedges and a thickness of $\qty{2.05}{\mm}$ close to the one of the wedges in the middle of the scan range. This leads to a usable scan range from -$\qty{15}{\fs}$ to +$\qty{25}{\fs}$.
Second, for IR pulse durations of $\qty{30}{\fs}$, a scanning range of typically -$\qty{45}{\fs}$ to +$\qty{45}{\fs}$ is required. To this end, another QP (QP2 in Fig.~\ref{fig:Wedges}) of thickness $\qty{1.03}{\mm}$ is added to extend the delay between the two pulses. It can be used in three configurations: placed after QP1 with its optical axis orientated vertically to scan the range -$\qty{50}{\fs}$ to -$\qty{10}{\fs}$ or horizontally to scan the range +$\qty{20}{\fs}$ to +$\qty{60}{\fs}$. To scan the range -$\qty{15}{\fs}$ to +$\qty{25}{\fs}$, it is placed before the wedges with its optical axis at the angle $\theta$, so that the total dispersion of the setup is preserved. By connecting the three scans during the analysis, we can access a total delay range of -$\qty{50}{\fs}$ to +$\qty{60}{\fs}$.
Third, the change in the thickness of the quartz during the scan induces a variable chirp in the IR pulse that modifies the pulse duration. To prevent this, we add two non-birefringent fused silica wedges (depicted by CW in Fig.~\ref{fig:Wedges}) whose variable thickness is adjusted in the opposite direction to the quartz wedges, compensating for the variation of the quartz thickness. Since quartz and silica have a similar group velocity dispersion (GVD) at $\qty{800}{\nm}$, the resulting chirp variation is limited to $\pm \qty{4}{\fs^2}$ during the scan. This ensures that the IR pulse profile is not significantly affected by the dispersion. % by precompensating the dispersion of the total glass thickness (around $\qty{4}{\mm}$).  

Before the HHG vacuum chamber an iris with a diameter of $\qty{5.2}{\mm}$ was used to reduce the pulse energy to $\qty{520}{\micro\J}$, which is then focused by a lens with $\qty{30}{\cm}$ focal length. This leads to a beam waist radius in the generating medium of $\qty{40}{\micro\meter}$ and an estimated peak intensity of $I_0 = \qty{3.4e14}{\W\per\square\cm}$ at focus. Harmonics are generated in a $\qty{6}{\mm}$ long gas cell filled with Ar at $\qty{4}{\milli\bar}$ backing pressure and analysed with a flat field spectrometer equipped with a $\qty{150}{\micro\m}$ wide entrance slit and an imaging grating with 300~gr/mm. This allows us to observe both the spatial and the spectral profiles of the harmonics simultaneously. %For the next results, we selected the spectra associated to central part of the beam.

\subsection{Data processing procedure}

\begin{figure*}[t]
    \centering
    \includegraphics[width=0.9\linewidth]{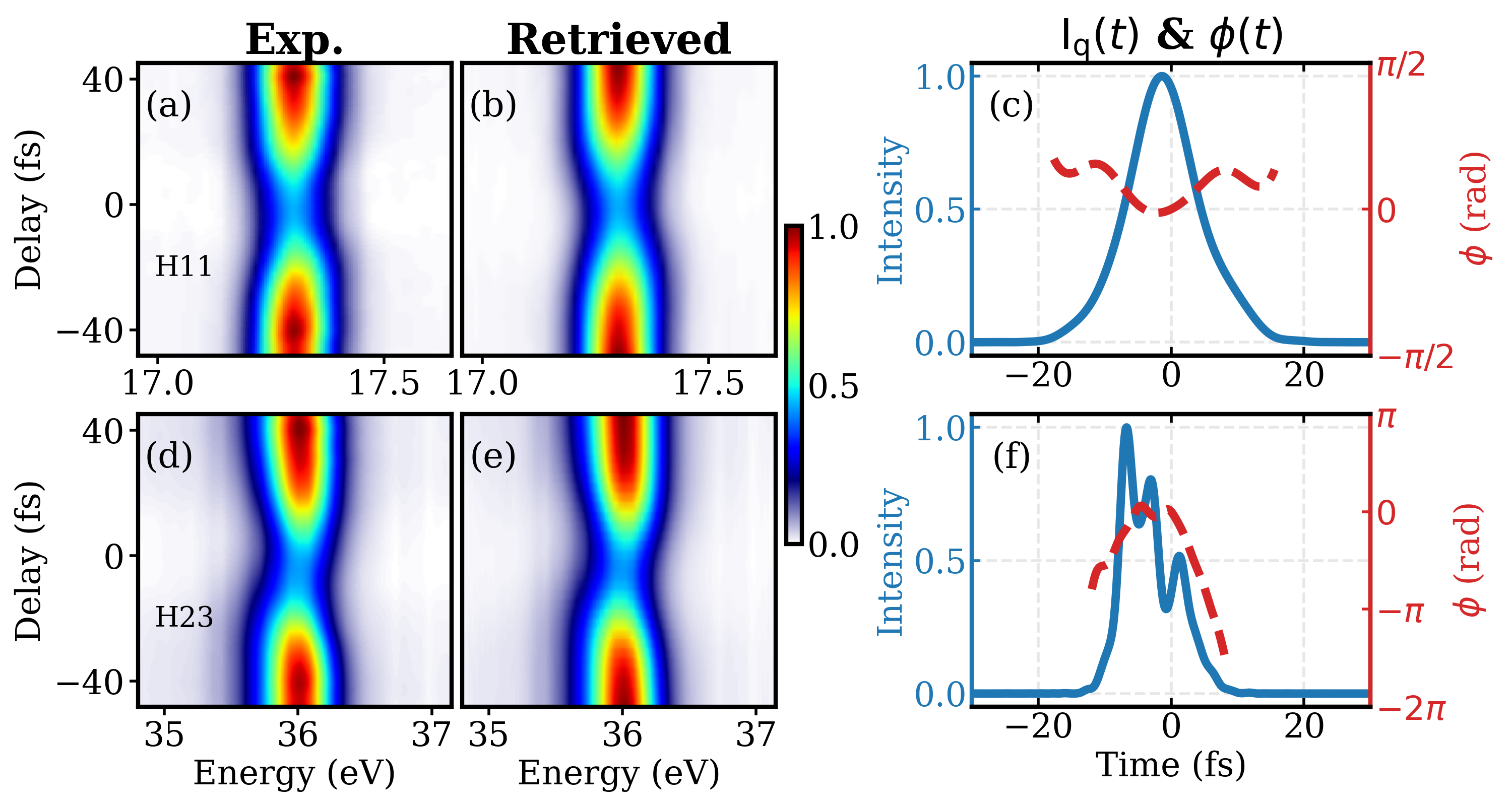}
    \caption{Reconstruction of experimental ellipticity perturbation traces for harmonic 11 (a) and harmonic 23 (d). Column 2: corresponding retrieved traces (b, e). Column 3: retrieved intensities (blue solid lines) and phases (red dashed lines).}
    \label{fig:exp_reconstructions_exemples}
\end{figure*}

Here, we describe the procedure for obtaining experimental traces that are afterwards analysed using the ptychographic algorithm. The scan was performed by translating the wedges in $\qty{50}{\micro\m}$ steps, which corresponds to $\qty{60}{\as}$ steps of delay between the driving pulse and the perturbation pulse. The IR pulse duration was measured to $\qty{29}{\fs}$ using a d-scan method \cite{mirandaSimultaneousCompressionCharacterization2012a,loriotSelfreferencedCharacterizationFemtosecond2013} and great care was taken to ensure that the IR pulse is not chirped in the interaction region. For each delay, the XUV spectrum is obtained by integrating the central part of the spatial profile in the far field. %add the distance to the detector? 
First, the signal is normalised after the full acquisition to account for any intensity fluctuations. 
%The first aspect we have to consider is the normalisation of the signal after the full acquisition. 
As presented in Section~\ref{sec:ellipt_theory}, the signal recorded for each harmonic manifests oscillations with a period of $T_0/2$. At the maxima of this signal ($\Delta \phi = 0$), the pump ellipticity is zero and the signal should be the same as the unperturbed one. We used this property to normalise the spectra by interpolating a curve passing through the experimental maxima (called the upper envelope) and dividing the signal by this curve. %By doing so, we eliminate slow fluctuations in the signal that can be attributed to laser fluctuations. % or small misalignments during the motion of the wedges. 
Second, it is necessary to combine the three scans performed in a single signal. With QP1 as used in our experiments, there is an overlap of $\qty{5}{\fs}$ between two consecutive scans, corresponding to three oscillations. By taking the average value of the two signals inside this overlap, we are able to smoothly connect the two scans. Fourier transform filtering is then applied to the complete signal to remove high-frequency noise. 
The last aspect is the oscillations. As mentioned in Section~\ref{sec:ellipt_theory}, we suppress the rapid oscillations of the signal due to the change in dephasing between main and perturbation pulses. For this purpose, we only keep the specific XUV spectra corresponding to the minima of the signal ($\Delta \phi = \pi/2$). Spectra mimicking a $\pi/2$ dephasing are then interpolated for all intermediate delays. 
In this way, we obtain the full ellipticity perturbation trace and can choose the number of delays and the delay range for the ptychographic analysis. %à retravailler une fois de plus 

\subsection{Experimental reconstruction}

We show the experimental traces for harmonics 11 and 23 in  Fig.~\ref{fig:exp_reconstructions_exemples}~(a) (d), respectively. These traces are qualitatively similar to the simulated ones. They differ significantly from each other, especially in terms of spectral width and response to ellipticity. For both traces, the spectra in the edges are invariant with respect to the delay, which means that the delay range is sufficiently large to access the unperturbed spectra. From the values of the lower envelopes at zero delay, we extract the experimental values for $\varepsilon_{\textrm{thr}}$ (given in Table~\ref{tab:eps_thr} with a constant error bar of $\pm 0.002$). %that decrease with the harmonic order and 
It is consistent with values reported in the literature \cite{mollerDependenceHighorderharmonicgenerationYield2012c,larsenSubcycleIonizationDynamics2016}. As seen in Fig.~\ref{fig:exp_reconstructions_exemples}~(d), the impact of the perturbation is stronger for negative delays, i.e.\ when the perturbation is in the rising front, unveiling that the XUV emission occurs mainly in this front. In this trace for harmonic 23, the "s-shape" of the central frequency is also clearly visible, which implies that the XUV pulse is chirped. Blue frequencies are more affected when the perturbation is in the rising front, which implies that they are emitted there. %, denoting the effect of a strong positive chirp (see \cref{sec:ellipt_theory}). 
Both of these effects are less visible for harmonic 11 in (a), which shows a quasi symetric evolution.

\begin{table*}[b]
    \centering
    \arrayrulecolor{black} 
    \begin{tabular}{c|ccccccccc}
    \toprule
       Harmonic order  & 11 & 13 & 15 & 17 & 19 & 21 & 23 & 25 & 27 \\ \hline
       $\varepsilon_{\textrm{thr}} (\pm 0.002)$  & $0.129$ & $0.124$ & $0.123$ & $0.123$ & $0.130$ & $0.130$ & $0.127$ & $0.124$ & $0.119$ \\
       \bottomrule 
    \end{tabular}
    \caption{Table of measured ellipticity threshold  $\varepsilon_{\textrm{thr}}$ for each harmonic order using the value of the lower envelope at zero delay.}
    \label{tab:eps_thr}
\end{table*}

The ellipticity perturbation traces retrieved for harmonics 11 and 23 (d, e) are presented in Fig.~\ref{fig:exp_reconstructions_exemples}~(b) and (e), respectively, and are in good agreement with the experimental ones (RMSE ~ 1\%). The corresponding retrieved XUV intensity (blue solid lines) and temporal phase (red dashed lines) profiles are displayed in panels (c) and (f). The features visible on the traces and discussed above are well reproduced by the retrieval algorithm. 
While harmonic 11 is emitted during the full pulse with a small chirp, harmonic 23 is mainly emitted in the rising front with a stronger chirp. It is interesting to note that an asymmetry in the temporal profile of H23 is clearly visible, with a sharp rise and a longer decay. The durations are also quite different, with the FWHM duration of H11 being $\qty{15}{\fs}$ FWHM, approximately half the duration of the IR pulse despite the highly nonlinear phenomenon, while H23 exhibits a duration of approximately $\qty{8}{\fs}$. 

%\textcolor{red}{ -- Add a section on the SFA model --}

Regarding the phases, since the instantaneous frequency is the derivative of the temporal phase, the higher frequencies in the rising pulse front indicate a positive chirp (see Section~\ref{sec:ellipt_theory}). This is confirmed by the phase profile $\phi(t)$ retrieved for H23. Moreover, the amplitude of the "s-shaped" central frequency directly gives the amplitude of the frequency variation and thus the curvature of $\phi(t)$. Interestingly, a chirp with opposite sign is retrieved for H11. This will be addressed in the next section, where we consider all harmonics obtained in the same experimental data set to appreciate the evolution of these observed features.

\section{Discussion}

\begin{figure}[t]
    \centering
    \includegraphics[width=1\linewidth]{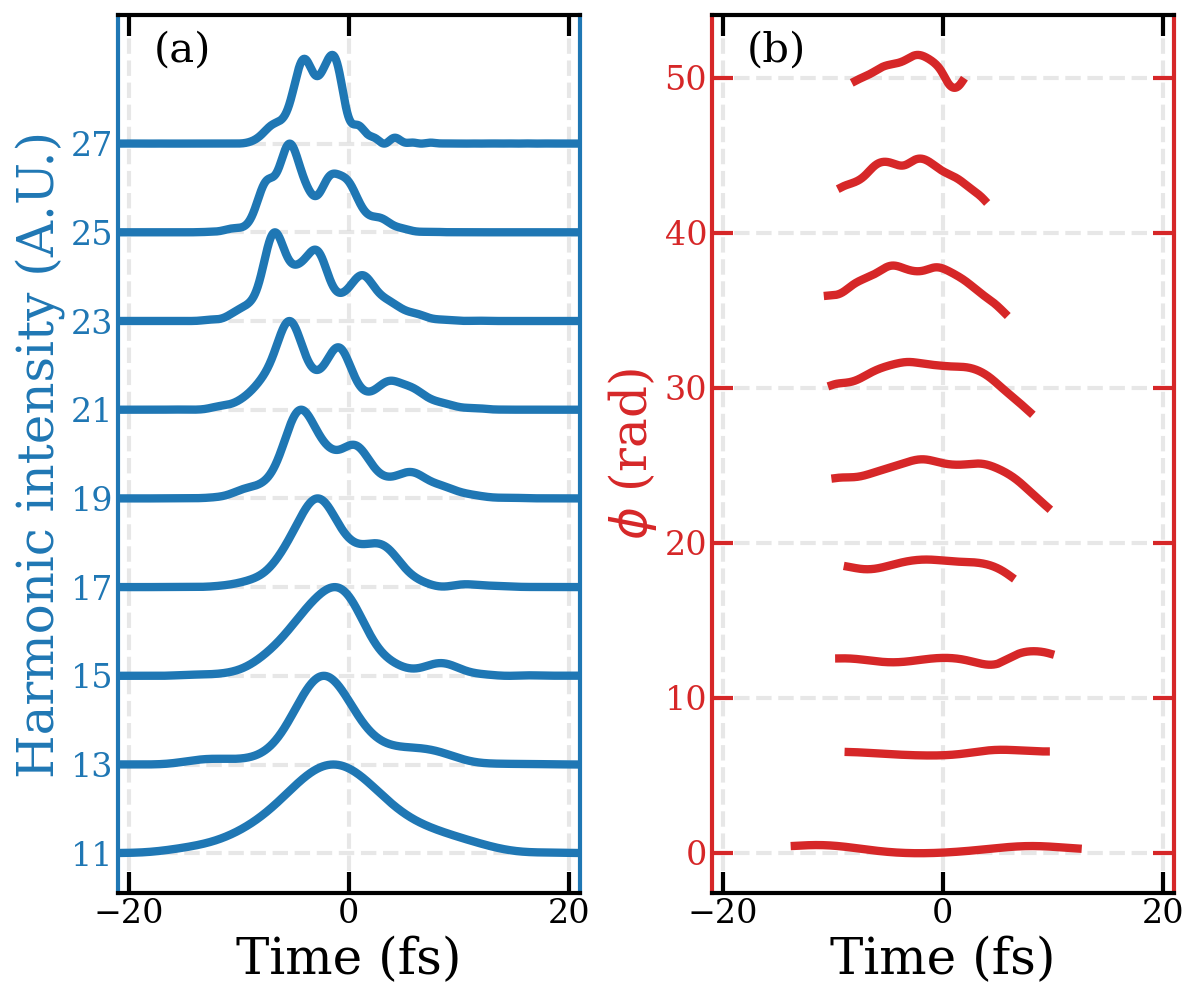}
    \caption{Waterfall plots of intensity (a) and phase (b) profiles retrieved for all harmonics from experimental ellipticity perturbation traces. The intensities are normalised and the numbers on the axis correspond to the harmonic order. The phases are only displayed in the time range where the intensity is higher than 10\% of the maximum intensity.}
    \label{fig:intensities_phases_reconstructed}
\end{figure}

The normalised retrieved intensity profiles for harmonic order from 11 to 27 are displayed in Fig.~\ref{fig:intensities_phases_reconstructed}~(a) with their corresponding temporal phases in (b). 
Looking first at the duration of the intensity profiles, there is a clear difference between harmonic 11, which has a long duration of around 15~fs, and the other harmonics, which have FWHM durations of less than 10~fs. 
H11 is also the only harmonic emitted throughout the entire IR pulse, while higher harmonics are mainly emitted in the rising front. This may be due to ionisation effects and time-dependent phase-matching that decrease the generation efficiency in the falling front of the IR pulse for those harmonics that are most sensitive to it \cite{kazamiasPressureinducedPhaseMatching2011, mevelOptimizingHighOrder2000}. 
We observe that the highest harmonics are emitted even before the maximum of the IR driving pulse is reached, which is compatible with a strong ionisation of the medium in accordance with the rather high pump intensity used. 
Another striking observation is the appearance of periodic structures inside the temporal profile of the highest order harmonics. The oscillation period seems to decrease with increasing harmonic order, leading to a higher number of peaks. 
The position of the main peak of the harmonic profile also evolves continuously with the harmonic order, further away from the centre of the IR pulse. 
%This behaviour is in an opposite way of what would be expected by only considering the single atom response \cite{corkumPlasmaPerspectiveStrong1993} where higher harmonics are emitted closer to the maximal intensity. This statement is true if we consider that the harmonic 27 is the last harmonic emitted but regarding the focused intensity in our generating medium the cutoff law gives us a maximal harmonic of around 50, which is effectively limited to the harmonic 35 by the saturation intensity in Ar. 
%That means that all the harmonics we are observing belong to the plateau. 
We attribute these evolutions to collective effects, such as phase-matching or spatiotemporal couplings. The coherence length depends both on the harmonic order $q$ and the intensity profile of the IR laser \cite{salieresCoherenceControlHighOrder1995a, rundquistPhaseMatchedGenerationCoherent1998, constantOptimizingHighHarmonic1999, kazamiasPressureinducedPhaseMatching2011} and plays a key role at high peak intensities, as is the case here. This results in a varying HHG efficiency within the IR pulse, which may create these evenly spaced peaks. Spatiotemporal couplings in the generating medium can effectively reshape the generating volume during the interaction, as the gas is ionised in the centre of the beam and the emission occurs in an annular domain \cite{dubrouilSpatioSpectralStructures2014}. Further investigations are necessary to study the relative effects of phase matching and ionisation. % at different pressures.% and with different spatial integrations of the XUV beam would be interesting to further study these hypothesis.

\begin{table*}[b]
    \centering
    \arrayrulecolor{black} 
    \begin{tabular}{c|ccccccccc}
    \toprule
       Harmonic order  & 11 & 13 & 15 & 17 & 19 & 21 & 23 & 25 & 27 \\ \hline
       $\alpha$ (exp.)  & $-0.7 \pm 0.4$ & $-0.7\pm 0.3$ & $0.1 \pm 0.2$ & $2.5 \pm 0.8$ & $3.8 \pm 1.2$ & $4.1 \pm 0.9$ & $4.7 \pm 1.5$ & $7.3 \pm 2.1$ & $12 \pm 2 $\\
       %$\alpha $ (SFA)  & $ -0.44 $ & $ -0.2 $ & $ -0.011 $ & $ 0.056 $ & $ 0.26 $ & $ 0.52 $ & $ 0.745 $ & $ 0.95 $ & $ 1.3 $\\
       %$\alpha $ (SFA)  & $ -1.19 $ & $ -0.73 $ & $ -0.11 $ & $ 0.64 $ & $ 1.66 $ & $ 2.85 $ & $ 4.52 $ & $ 6.35 $ & $ 11.6 $\\
       \bottomrule 
    \end{tabular}
    \caption{Table of estimated values of $\alpha$ in units of $10^{-14} \,  \qty{}{W^{-1} \, cm^2}$ for each harmonic order. The values obtained from experiments in Ar extracted using HORNET are presented  %in the first row 
    assuming a peak intensity of $I_0 = \qty{3.4e14}{\W\per\square\cm}$. %  
    }
    \label{tab:alpha_values}
\end{table*}

Second, the XUV temporal phases show a steady evolution with the harmonic order. The phase profiles can be approximated by quadratic functions whose radii of curvature increase with the harmonic order, implying an increasing chirp. This evolution of the harmonic chirp is consistent with the dipole model for the short trajectory. Considering a Gaussian profile for the pump intensity $I_{\text{IR}}(t)$, the temporal phase $\phi(t) = \alpha I_{\text{IR}}(t)$ can be approximated by a quadratic function around its maximum value. The SFA model and TDSE \cite{balcouQuantumpathAnalysisPhase1999, gaardeQuantumPathDistributions2002, varjuFrequencyChirpHarmonic2005} predict that $\alpha$ values for the short trajectory increase with the harmonic order, producing a larger femtochirp as observed in our measurements. This is consistent with the assumption that the short trajectory is predominant in the centre of the beam. Interestingly, the chirp for H11 and H13 is negative: the parabola have an opposite direction compared to the parabola of higher harmonics. This corresponds to negative $\alpha$ values as predicted by the SFA model \cite{varjuFrequencyChirpHarmonic2005, heSpatialSpectralProperties2009, yostVacuumultravioletFrequencyCombs2009, powerXFROGPhaseMeasurement2010} for lower harmonics and observed in previous experimental measurements \cite{carlstromSpatiallySpectrallyResolved2016}. The fitted values of $\alpha$ (which are assumed constant) are indicated in Table~\ref{tab:alpha_values} together with their uncertainties. Comparison can not be quantitative as we measure the collective response while SFA considers the single atom response. Nevertheless, the values obtained are of the same order of magnitude as the values obtained from SFA \cite{heSpatialSpectralProperties2009, varjuFrequencyChirpHarmonic2005}. 

The correct retrieval of the harmonic phase evolution further validates our approach. Note, however, that this method provides no link between the phases of individual harmonics and the attosecond temporal profile can not be reconstructed, yet. Further investigations will be performed using multiple overlapping harmonics to reconstruct the complete temporal profile of the attosecond pulse train. 

%mention of the THz method [Ardana-Lamas 2016] ? mention HHSPIDER ? A réécrire cette partie sur alpha effectif 

%The next step for this method will be to reconstruct the complete attosecond pulse train. For now, a phase relation is missing between consecutive harmonics, which prevents us to sum all the fields and reconstruct the attosecond pulse train. This lack of information can be circumvented in the case of overlapping harmonics. % but further investigations will provide the required . %Further investigations will provide the  the ability of the algorithm to reconstruct the electric field of two consecutive harmonics on condition that they overlap because it provides a reconstructed continuous spectral phase over the entire spectrum. 

\section{Conclusion}

We have developed an \textit{in situ} method for measuring the time dependent electric fields of individual harmonics produced by HHG with a linearly polarised femtosecond laser. Our approach exploits the strong sensitivity of HHG on the driving field ellipticity. 
By introducing time-dependent ellipticity in the driving IR field, the XUV generation is perturbed at specific times. These ellipticity perturbations are then used to probe the unperturbed XUV emission, and XUV spectra are recorded for several perturbation patterns. This provides the so-called ellipticity perturbation traces.
We have presented the technique, its theoretical description, and its experimental requirements. We have developed a ptychographic algorithm to retrieve harmonic electric fields with complex temporal profiles from the ellipticity perturbation traces. 
We validated and confirmed the robustness of the algorithm to noise and uncertainties in the IR temporal profile by using simulated ellipticity perturbation traces for which the harmonic electric fields are known. Finally, we experimentally measured perturbed XUV spectra and applied our technique and analysis to experimental traces to reconstruct the electric fields for several harmonics. We observed characteristic features of the generation, such as the XUV pulse durations between half and a quarter of the IR pulse duration. We also extracted the temporal phase of the XUV emission and observed a regular evolution of the femtochirp in agreement with predicitions from single atom SFA theory. % without any free parameter.  

This all-optical method relies only on measuring HHG spectra and therefore requires only basic equipment (XUV spectrometer and polarisation-splitting interferometer). It is a compact method that is easy to implement. Using birefringent materials yields a high resolution and intrinsically stable interferometer that is very convenient for this type of measurements. Furthermore, ptychographic reconstruction turns out to be a very robust technique for this unique and powerful retrieval approach of high-order harmonic temporal profiles. This technique is thus highly suitable for experimentally measuring the time-dependent electric field profiles of individual high-order harmonics \textit{in situ}. \newline

%\newpage

%\runin{Funding.} 

\runin{Acknowledgment.} The authors acknowledge Anne L'Huillier, Cord Arnold, Franck Lepine, Vincent Loriot, Emilien Prost and Eric Mevel for fruitful discussions.

\runin{Disclosures.} The authors declare no conflicts of interest.

\runin{Data availability.} The data that supports the finding of this study are available from the corresponding authors upon reasonable request.

%\runin{Supplemental document.} See Supplement 1 for supporting content. 

\bibliography{Article_HORNET}
\end{document}